\documentclass[12pt,draftclsnofoot, perreview, onecolumn]{IEEEtran}
\IEEEoverridecommandlockouts
\usepackage[utf8]{inputenc}
\usepackage[T1]{fontenc}
\usepackage{url}
\usepackage{ifthen}
\usepackage{cite}
\usepackage{amsmath} 
\usepackage{array}
\usepackage{nicematrix}                          
\usepackage{amsthm}
\usepackage{amsfonts}
\usepackage{mathrsfs}
\usepackage{multirow}
\usepackage{citesort}
\usepackage[caption=false]{subfig}
\usepackage{bm}
\usepackage{algorithm}
\usepackage{graphicx}
\usepackage{epstopdf}
\usepackage{amssymb}
\usepackage{diagbox}
\usepackage{makecell}
\usepackage{bigstrut}
\usepackage{tikz}

\usepackage{algpseudocodex}
\usetikzlibrary{patterns}
\usetikzlibrary{matrix,decorations.pathreplacing}
\usetikzlibrary{decorations.pathreplacing,calligraphy}

\usepackage{xhfill}
\allowdisplaybreaks[4]
\newcommand{\xfilll}[2][1ex]{%
	\dimen0=#2\advance\dimen0 by #1%
	\leaders\hrule height \dimen0 depth -#1\hfill%
}

\newenvironment{Proof}{
	{\indent\it Proof:}\;}{\hfill $\blacksquare$\par}
\newtheoremstyle{thry}
{3pt}
{3pt}
{}
{1em}
{}
{:}
{.5em}
{}
\theoremstyle{thry}

\newtheorem{theorem}{\emph{\textbf{Theorem}}}

\newtheorem{example}{\emph{Example}}

\ifCLASSOPTIONonecolumn
\newcommand{\figwidth}{0.65\textwidth}

\fi

\ifCLASSOPTIONtwocolumn
\newcommand{\figwidth}{0.48\textwidth}

\fi

\begin{document}
\title{Random Staircase Generator Matrix Codes: Performance Analysis and Construction \\
\thanks{This work was supported by the National Key R\&D Program of China~(Grant No.~2021YFA1000500) and the NSF of China~(No.~61971454 and No.~62301617). Part of this work was presented at the 2024 IEEE International Symposium on Information Theory~\cite{Wang2024ISIT}. \emph{(Corresponding author: Xiao Ma.)} }
}

\author{
  \IEEEauthorblockN{Qianfan Wang,~\IEEEmembership{Member,~IEEE}, Yiwen Wang, Yixin Wang, Jifan Liang\\ and Xiao Ma,~\IEEEmembership{Member,~IEEE} }\\

 \thanks{Qianfan Wang, Yiwen Wang, Jifan Liang and  Xiao Ma are with the School of Computer Science and Engineering and Guangdong Key Laboratory of Information Security Technology, Sun Yat-sen University, Guangzhou 510006, China~(e-mail:wangqf26@mail.sysu.edu.cn; wangyw93@mail2.sysu.edu.cn; liangjf56@mail2.sysu.edu.cn; maxiao@mail.sysu.edu.cn)}
 \thanks{ Yixin Wang is with the School of Systems Science and Engineering, Sun Yat-sen University,  Guangzhou 510006, China~(e-mail:
 	wangyx58@mail2.sysu.edu.cn)}
}

\maketitle

\begin{abstract}
In this paper, we propose a class of codes, referred to as random staircase generator
matrix codes~(SGMCs), which have staircase-like generator matrices.
In the infinite-length region, we prove that the random SGMC is capacity-achieving over binary-input output-symmetric~(BIOS) channels.
In the finite-length region, we present the representative ordered statistics decoding with local constraints~(LC-ROSD) algorithm for the SGMCs.
The most distinguished feature of the SGMCs with LC-ROSD is that the staircase-like matrices enable parallel implementation of the Gaussian elimination~(GE), avoiding the serial GE of conventional OSD and supporting a potential low decoding latency, as implied from simulations.
To analyze the performance of random SGMCs in the finite-length region, we derive the ensemble weight spectrum and invoke the conventional union bound. We also derive a partially random coding union~(RCU) bound, which is tighter than the conventional one and is used as a criterion to design the SGMCs.
Staircase-like generator matrices allow us to derive a series of~(tighter and tighter) lower bounds based on the second-order Bonferroni inequality with the incremental number of codewords.
The numerical results show that the decoding performance can match well with the proposed partially RCU bound for different code rates and different profiles.
The numerical results also show that the tailored SGMCs with the LC-ROSD algorithm can approach the finite-length performance bound, outperforming the 5G low-density parity-check~(LDPC) codes, 5G polar codes, and Reed-Muller~(RM) codes.
\end{abstract}


\section{Introduction}

The short block error-correction codes are crucial for the ultra-reliable low-latency communication~(URLLC) in 5G~\cite{5GURLLC} and the hyper-reliable and low-latency communication~(HRLLC) in 6G~\cite{union2023future}.
It has been reported in~\cite{cocskun2019efficient} that several existing short codes can approach the random coding union~(RCU) bounds by using the near-maximum-likelihood (ML) decoding algorithms, which however, usually have high decoding complexity and high decoding latency.

The ordered statistics decoding~(OSD)~\cite{fossorier1995soft} is a typical near-ML decoding algorithm\footnote{It has been proved that OSD with a tailored test order and a sufficient stopping criterion is an ML algorithm~\cite{Ma2024}.}.
The OSD is also universal in the sense that it can be used to decode any short linear block codes, including Bose-Chaudhuri-Hocquenghem~(BCH) codes, low-density parity-check (LDPC) codes, turbo codes, and polar codes, resulting in capacity-approaching performance~\cite{Park2019,jiang2007reliability,wei2013crc,wu2016ordered}.
The underlying idea of the original OSD is that the errors tend to occur less frequently in the most reliable positions compared to the least reliable positions.
As a result, by flipping a small number of bits~(with the maximum number known as order $t$) in the most reliable basis~(MRB) and re-encoding, a list of candidate codewords can be generated, from which the most likely one is selected as the decoding output.
To approach the ML decoding performance, it has been proved that, for a linear block code with minimum Hamming distance $d_{\min}$, the order of $t=\left\lceil d_\text{min}/4-1\right\rceil$ is required~\cite{fossorier1995soft}.
One issue of the original OSD is that the decoding with order~$t$ requires generating and re-encoding of $\sum_{i=0}^t \binom{k}{i}$ test error patterns (TEPs), resulting in a computational complexity of order $\mathcal{O}(k^t)$.
Another issue is the Gaussian elimination~(GE) used in the conventional OSD, which has a complexity of order $\mathcal{O}(k^3)$.
Even worse, in the conventional OSD, GE can only be implemented in serial~(column by column) for general matrices, leading to an additional decoding latency.

To reduce the complexity arising from the re-encoding process in OSD, several improved OSD algorithms have been proposed, including the segmentation-discarding OSD~(SD-OSD)~\cite{Yue2019SD}, the linear-equation OSD~(LE-OSD)~\cite{yue2022linear}, and the probability-based OSD~(PB-OSD)~\cite{Yue2021PB}, among others.
Recently, the OSD with local constraints (LC-OSD) has been introduced in~\cite{Wang2022LCOSD,Liang2023LCOSD} and applied in joint source-channel coding~\cite{Wang2024TCOMJSCC,Wang2023Globecom,Chen2023WCSP}.
Unlike the conventional OSD that selects the most reliable basis (MRB) of dimension $k$, the LC-OSD selects $k + \delta$ bits as an extended MRB, introducing local constraints on the TEPs.
This allows the OSD to skip numerous invalid TEPs.
By optimizing the test order with the tailored early stopping criteria, the average number of TEPs can be significantly reduced, even down to
less than ten from hundreds of thousands as required by the original OSD, in the high signal-to-noise ratio~(SNR) region.
Now, the issue caused by the serial GE becomes more critical for the practical implementation of the OSD.
This issue can be circumvented in some cases by precalculating and storing multiple systematic generator matrices~\cite{choi2021fast} or by invoking specific decoding conditions to skip the GE process~\cite{yue2022ordered}.
If the constraint on the MRB is relaxed, the GE can be replaced by the polynomial interpolations to form an enlarged basis over an extension field for BCH codes~\cite{yang2022low} or a near reliable basis~\cite{QuasiOSD} for binary images of RS codes.

In contrast to most existing works that aim to simplify OSD for general codes,
this paper introduces a new class of codes~(referred to as random SGMCs) that are specifically tailored for efficient decoding using a low-complexity OSD-like algorithm.
The proposed random SGMCs are capacity-achieving~(demonstrated by theoretical proof) in the asymptotic regime as the code length tends to infinity and are capacity-approaching~(demonstrated by the performance bounds and the numerical results) in the finite code length regime.
The main contributions of this paper are summarized as follows.

\begin{enumerate}
  \item \textbf{New Code Ensemble}: We propose a class of codes, referred to as random staircase generator matrix codes~(SGMCs), which have staircase-like random generator matrices.
      Specifically, the random staircase matrix is a lower triangular matrix with ones on the staircase, zeros on the upper right to the staircase, and random elements on the lower left to the staircase.

  \item \textbf{Coding Theorem}: In the infinite-length region, we prove that the random SGMC is capacity-achieving over the binary-input output-symmetric~(BIOS) channels.

  \item \textbf{Decoding Algorithm}:
   For practical use, we turn to the representative OSD~with local constraints~(LC-ROSD) algorithm for the SGMCs.
   Unlike the original OSD, which selects $k$ reliable bits to form the MRB, the proposed LC-ROSD algorithm selects $k+\delta$ bits~(at least one from each staircase) to form an extended relatively reliable basis, introducing the local constraints.
   The order of the TEP can be optimized by using the serial list Viterbi algorithm~(SLVA) or the two-way flipping pattern tree~(FPT) algorithm over the local constraints.
   More importantly, the staircase-like matrices enable parallel implementation of the GE, avoiding the serial GE of conventional OSD and supporting a potentially low decoding latency, as implied from the numerical results.

  \item \textbf{Performance Analysis}:
  To analyze the performance of random SGMCs in the finite-length region, we derive the ensemble weight spectrum and invoke the conventional union bound.
  We also derive a partially RCU bound, which is tighter than the conventional one and well-matched to the simulation results. Staircase-like generator matrices allow us to derive a series of~(tighter and tighter) lower bounds based on the second-order Bonferroni inequality with the incremental number of codewords.

  \item \textbf{Construction Rules}: Given the proposed partially RCU bound, we propose a criterion to design the SGMCs, where the profile is optimized to achieve a trade-off between performance and decoding complexity.



  \item \textbf{Numerical Results}: The numerical results show that the random SGMCs with the LC-ROSD algorithm can match well with the proposed partially RCU bound for different code rates and different profiles.
      The numerical results also show that the tailored SGMCs with the LC-ROSD algorithm can approach the finite-length performance bound, outperforming the 5G LDPC codes, 5G polar codes, and Reed-Muller~(RM) codes.
\end{enumerate}

The rest of this paper is organized as follows. In Section~\ref{sec2}, we present the system model and the definition of the SGMCs. We also prove that the random SGMC is capacity-achieving in the infinite-length region by using the derived error exponent in this section.
In Section~\ref{sec3}, we focus on the finite-length random SGMCs.
We present the LC-ROSD algorithm of the SGMCs for practical use.
The SGMCs are analyzed by deriving the ensemble weight spectrum, the partially RCU bounds, and the second-order Bonferroni lower bounds.
The code design for the profile and the performance comparison are presented in  Section~\ref{sec4}.
Section~\ref{sec5} concludes this paper.

Notation: Denote by $\log$ the base-2 logarithm and by $\exp$ the base-2 exponent. Denote by upper-case letter, say $X$, a random variable and
denote by the corresponding lower-case letter $x \in \mathcal{X}$ the realization of the random variable.
We represent a vector of length $m$ as $\bm{x} = (x_0, x_1, \cdots, x_{m-1})$.
To emphasize the length of $\bm{x}$, we also use the notation of  $\bm{x}^m$.
We use $P_X(x), x \in \mathcal{X}$  to denote the probability density function~(PDF) of a continuous random variable or the probability mass function~(PMF) of a discrete random variable.
If the context is clear, we may omit the subscript of the PDF or the PMF.
We use $P^n(x)$ to denote the probability function obtained by independently replicating the probability distribution $P(x)$ for $n$ times.

\section{Staircase Generator Matrix Codes}\label{sec2}

\subsection{System Model}
Let $\mathbb{F}_2 = \{0,1\}$ be the binary field. Let $\mathscr{C}[n,k]$  be a binary linear block code of length $n$ and dimension $k$, whose generator matrix and parity-check matrix are denoted by $\mathbf{G}$ and $\mathbf{H}$, respectively.
At the transmitter, the information sequence $\bm U^{k}\in \mathbb{F}_2^{k}$ is firstly encoded as $\bm X^n \in \mathbb{F}_{2}^{n}$ by $\mathscr{C}[n,k]$, i.e., $\bm X^n=\bm U^k{\textbf{G}}$.
Then, the codeword $\bm X^n$ is transmitted through a BIOS memoryless channel, resulting in $\bm Y^n$.
At the receiver, the decoder of $\mathscr{C}[n,k]$ takes as input the receiving sequence $\bm Y^n$ and delivers as output the estimated information sequence $\hat{\bm U}^k$.
The performance of the system model is measured by FER, which is defined as ${\text{FER}} = {\rm Pr}\{\hat{\bm U}^k\neq \bm U^k\}$.

The BIOS memoryless channel is characterized by an input $X \in \mathcal{X} = \mathbb{F}_{2}$, an output set $\mathcal{Y}$~(discrete or continuous), and a conditional probability mass (or density) function $\{P_{Y|X}(y|x)\big| x\in\mathbb{F}_2, y\in\mathcal{Y}\}$ which satisfies the symmetric condition that $P_{Y|X}(y|1)=P_{Y|X}(\pi(y)|0)$ for some invertible mapping $\pi~: \mathcal{Y}\rightarrow\mathcal{Y}$ with $\pi^{-1}(\pi(y))=y$.
The channel is said to be memoryless if $P_{\bm{Y}|\bm{X}}(\bm y|\bm x)=\prod\limits_{i=0}^{n-1}P_{Y|X}(y_i|x_i)$.
For a Bernoulli input $X$, we can define the mutual information $I(X; Y)$.
The channel capacity of the BIOS channel is given by $C_{\rm BIOS} = I(X; Y)$ with $X$ being a uniform binary random variable with $P_X(0) = P_X(1) = 1/2$.

\subsection{Staircase Generator Matrix Codes}
\textbf{Definition 1~(Staircase Generator Matrix Codes)}.
Let $w_\ell, 0\leq \ell \leq k-1$, be $k$ positive integers such that $\sum_{\ell = 0}^{k-1} w_\ell = n$.
A binary linear block code $\mathscr{C} [n, k]$ is referred to as a staircase generator matrix code if it has a generator matrix with the $\ell$-th row specified by
$\bm{g}_\ell = (\bm{h}_\ell, \bm{i}_\ell, \bm{j}_\ell)$, where $\bm{h}_\ell$ is a  binary subsequence of length $\sum_{i=0}^{\ell-1} w_i$, $\bm{i}_\ell$ is the all-one subsequence of length $w_\ell$, and  $\bm{j}_\ell$ is the all-zero subsequence of length $n - \sum_{i=0}^{\ell} w_i$.
Here, the $k$-tuple $\bm{w} = ({w}_0, {w}_1, \cdots, {w}_{k-1})$ is referred to as the profile of the staircase generator matrix and $w_\ell$ is the width of the $\ell$-th staircase $\bm{i}_\ell$.

It is not difficult to see that, possibly with column permutations, polar codes~\cite{2009Arikan} and RM codes~\cite{reed1954class,Muller1954} are SGMCs, as illustrated by the following example.
For convenience, we call a profile $\bm{w} = (w_0,w_1,\cdots,w_{k-1})$ as an RM-profile if it is the profile of some RM codes or polar codes.
\begin{example}
  The first-order RM code $\mathscr{C}[8,4]$ has a staircase generator matrix given by
\begin{equation}
  {\bf {G}}_{\text{RM}}=\begin{pNiceMatrix}%
  [margin,
  name=polar,
  nullify-dots,
  xdots/line-style=loosely dotted,
  code-after = {\begin{tikzpicture}[decoration={calligraphic brace, mirror, amplitude=6pt,raise=2pt}][matrix of nodes]
                  \draw[thick,red,-,yshift=0em] (-4.7,1.00)|-(-2.6,1.00)|-(-2.0,0.28)|-(-1.5,0.28)|-(-0.9,-0.45)|-(-0.4,-1.2);
                  \end{tikzpicture}}
  ]
   \textcolor{blue}1 & \textcolor{blue}1 & \textcolor{blue}1 & \textcolor{blue}1 & 0 & 0 & 0 &0\\
    1 & 1 & 0 & 0 & \textcolor{blue}1 & \textcolor{blue}1 & 0 &0\\
    1 & 0 & 1 & 0 & 1 & 0 & \textcolor{blue}1 &0\\
    1 & 1 & 1 & 1 & 1 & 1 & 1 &\textcolor{blue}1\\
  \end{pNiceMatrix},
  \label{RM-permutation}
\end{equation}
where the RM-profile is $\bm{w}={(4,2,1,1)}$.
In this example, the $\ell$-th staircase  $\bm{i}_\ell$ is presented as $w_\ell$ blue 1s and the staircase-like structure is visually depicted using red lines.
\end{example}

\textbf{Definition 2~(Random Staircase Generator Matrix Codes)}.
An SGMC is referred to as a random SGMC if the elements under the staircase~(i.e., $\bm{h}_\ell$ for all $\ell > 0$) are sampled from a Bernoulli process with success probability $1/2$.

\emph{{Example \rm{1}~(Cont’d)}}: The random SGMC with the RM-profile of length $n = 8$ and dimension $k = 4$ has the generator matrix given by
\begin{equation}
  {\bf {G}}_{\text{random SGMC}}=\begin{pNiceMatrix}%
  [margin,
  name=random SGMCs,
  nullify-dots,
  xdots/line-style=loosely dotted,
  code-after = {\begin{tikzpicture}[decoration={calligraphic brace, mirror, amplitude=6pt,raise=2pt}][matrix of nodes]
                  \draw[thick,red,-,yshift=0em] (-4.7,1.00)|-(-2.6,1.00)|-(-2.0,0.28)|-(-1.5,0.28)|-(-0.9,-0.45)|-(-0.4,-1.2);
                  \end{tikzpicture}}]
   \textcolor{blue}1 & \textcolor{blue}1 & \textcolor{blue}1 & \textcolor{blue}1 & 0 & 0 & 0 &0\\
    * & * & * & * & \textcolor{blue}1 & \textcolor{blue}1 & 0 &0\\
    * & * & * & * & * & * & \textcolor{blue}1 &0\\
    * & * & * & * & * & * & * &\textcolor{blue}1\\
  \end{pNiceMatrix},
  \label{RM-permutation-a}
\end{equation}
where the elements marked by * are sampled from a Bernoulli process with success probability $1/2$.

\subsection{Coding Theorem for Staircase Generator Matrix Codes}

\par By assuming that all codewords are randomly generated according to an identical distribution, the error exponent was derived in~\cite{Gallager1968} to prove the channel coding theorem, where the pair-wise independence between codewords is required.
This requirement can be circumvented by assuming that the all-zero codeword is transmitted and deriving the partial error exponent~\cite{Ma2022ISIT}, which serves as a powerful tool to prove the coding theorem for binary linear codes over the BIOS channels.

Let $P(1)=p$ and $P(0)=1-p$~be the input distribution for a BIOS memoryless channel. The mutual information between the input and the output is given by

\begin{equation}
	I(p)=(1-p)I_0(p)+pI_1(p),
\end{equation}
where
\begin{equation}
	I_0(p)=\sum\limits_{y\in\mathcal{Y}} P(y|0)\log\frac{P(y|0)}{P(y)},
	\label{I_{0}}
\end{equation}
\begin{equation}
	I_1(p)=\sum\limits_{y\in\mathcal{Y}} P(y|1)\log\frac{P(y|1)}{P(y)},
\end{equation}
and $P(y)=(1-p)P(y|0)+pP(y|1)$. We call $I_0(p)$ $\big({\rm or}~I_1(p)\big)$ \emph{partial mutual information}. For the BIOS memoryless channel, we have ${\rm max}_{0\leq p \leq 1}I(p)=I(1/2)=I_0(1/2)=I_1(1/2)$, which is the channel capacity. Notice that $I_0(p) > 0$ for $0<p<1$ as long as ${\rm Pr}\{y |P(y|0) \neq P(y|1)\} > 0$. This is a natural assumption in this paper. It can be proved directly that~\cite{Ma2022ISIT} the partial mutual information $I_0(p)$ is continuous, differentiable and increasing from $I_0(0)=0$ to $I_0(1)$.

\begin{theorem}\label{theorem_1}
Let $\bm{w} = (w_0,w_1,\cdots,w_{k-1})$ be a profile. The ensemble average FER over the random SGMCs specified by $\bm{w}$ can be bounded by
\begin{equation}
		\label{conditional_pro}
		{\rm FER}_{\text{avg}} \leq \sum_{\ell = 0}^{k-1}\exp\left[- n_\ell E(w_\ell,R_\ell) \right] ,
	\end{equation}
where $n_\ell = \sum_{i = 0}^{\ell} w_i$, $R_\ell = \ell / n_\ell$,
\begin{equation}
		E(w_\ell, R_\ell)=\max\limits_{0\leq\gamma\leq 1}\left( \frac{n_\ell - w_\ell}{n_\ell}E_0\left(\frac{1}{2},\gamma \right) + \frac{w_\ell}{n_\ell}E_0\left(1,\gamma\right) - \gamma R_\ell \right),
\label{Eq:E_ell}	
\end{equation}
and
	\begin{equation}
		\begin{aligned}
			E_0(p,\gamma)=-\log\Bigg\{\sum\limits_{ y\in \mathcal{Y}}P( y| 0)^{\frac{1}{1+\gamma}}&\Big[(1-p)(P( y| 0))^{\frac{1}{1+\gamma}}+p(P( y| 1))^{\frac{1}{1+\gamma}}\Big]^\gamma\Bigg\} \text{.}
		\end{aligned}	
	\end{equation}
\end{theorem}

\begin{Proof}
For BIOS memoryless channels, without loss of generality, suppose that $\bm{0} =  \bm{0} {\rm {\bf G}}$ is transmitted.
Upon receiving $\bm y$, we choose a sequence ${\bm u}$ such that $P(\bm y|{\bm u}{\rm {\bf G}} = {\bm c})$ is maximized.

To derive the coding theorem, we partition the nonzero messages into $k$ disjoint subsets, i.e., $\mathbb{F}_2^k \setminus \{\bm{0}\} = \bigcup_{\ell = 0}^{k-1}\mathcal{U}_\ell$, where the subset $\mathcal{U}_\ell$ for $\ell \geq 0$ consists of $2^\ell$ sequences whose right-most 1 appears at the position $\ell$. In other words, $\bm{u} \in \mathcal{U}_\ell$ if and only if $u_\ell = 1$ and $u_i = 0$ for all $i > \ell$.
For $0 \leq \ell \leq k-1$, we denote collectively by $\mathscr{C}_\ell$ all the $2^\ell$ codewords corresponding to the input information set $\mathcal{U}_{\ell}$, i.e., $\mathscr{C}_\ell = \{\bm c= \bm u {\bf G}|{\bm u} \in \mathcal{U}_\ell\}$.
Due to the randomness of the staircase generator matrix, the codewords in $\mathscr{C}_\ell~(0\leq \ell \leq k-1)$ are random vectors and share an identical distribution. Precisely, a codeword $\bm c \in \mathscr{C}_\ell$ can be divided into three segments, where the first segment of length $\sum_{i< \ell}w_i$ is sampled from a Bernoulli process of success probability $1/2$, the second segment of length $w_\ell$  is fixed to the all-one vector and the third segment of length $n - \sum_{i\leq \ell}w_i$ is fixed to the all-zero vector.

Given a received sequence $\bm y$, let $A_\ell$ be the event that some $\bm{u} \in \mathcal{U}_\ell$ exists such that $P(\bm y|{\bm u}{\bf G})\geq P(\bm y|\bm 0)$.
We have
	\begin{equation}
		\begin{aligned}
			{\rm Pr}\{{\text{error}}|\bm 0\} &=\sum\limits_{\bm y\in \mathcal{Y}^{n}}P(\bm y|\bm 0)\cdot {\rm {Pr}}\left\{\bigcup\limits_{\ell=0}^{k - 1}   A_{\ell}\right\}\\
 &\leq \sum\limits_{\ell = 0}^{k-1} \sum\limits_{\bm y\in \mathcal{Y}^{n}}P(\bm y|\bm 0)\cdot {\rm {Pr}}\left\{   A_{\ell}\right\}.
		\end{aligned}
		\label{P}	
	\end{equation}
Since $\bm c = \bm u \bf G$ for all $\bm u \in \mathcal{U}_\ell$ share a common distribution, from~\cite[Lemma in Chapter 5.6]{Gallager1968}, we have
	\begin{equation}
		\begin{aligned}
	&\sum\limits_{\bm y\in \mathcal{Y}^{n}}P(\bm y|\bm 0)\cdot {\rm {Pr}}\left\{ A_{\ell}\right\} \leq  2^{\ell\gamma} \sum\limits_{\bm y\in \mathcal{Y}^{n}}P(\bm y|\bm 0)\left({\rm {Pr}}\left\{P(\bm y|{\bm c})\geq P(\bm y|\bm 0)\right\}\right)^\gamma,
		\end{aligned}
		\label{P}	
	\end{equation}
for any given $0\leq \gamma\leq 1$.
From Markov inequality, for $s=1/(1+\gamma)$ and a given received vector $\bm y$, the probability of a codeword being more likely than $\bm 0$ is upper bounded by
\begin{equation}
		\begin{aligned}
			{\rm Pr}\{P(\bm y|{\bm{c}})\geq P(\bm y|\bm 0)\}&\leq  \frac{{  \rm{\bf E}}[(P(\bm y|{\bm{c}}))^{s} ]}{(P(\bm y|\bm 0))^{s}}\\
			&=\sum\limits_{{\bm{c}}} P({\bm{c}})\frac{(P(\bm y|{\bm{c}}))^{s}}{(P(\bm y|\bm 0))^{s}}\text{.}
		\end{aligned}
		\label{Markov}
\end{equation}
By substituting the bound of (\ref{Markov}) into \eqref{P}, for $\ell \geq 0$, $n_\ell = \sum_{i=0}^{\ell}w_i$ and $R_\ell = \ell/n_\ell$, we have
	\begin{equation}
		\begin{aligned}
 & \sum\limits_{\bm y\in \mathcal{Y}^{n}}P(\bm y|\bm 0)\cdot {\rm {Pr}}\left\{ A_{\ell}\right\}   \leq  2^{\ell\gamma} \sum\limits_{\bm y\in \mathcal{Y}^{n}}P(\bm y|\bm 0)\left({\rm {Pr}}\left\{P(\bm y|{\bm c})\geq P(\bm y|\bm 0)\right\}\right)^\gamma \\ &\leq 2^{\ell\gamma} \sum\limits_{\bm y\in \mathcal{Y}^{n_\ell}}P(\bm y|\bm 0) \left[\sum\limits_{{\bm{c}}} P({\bm{c}})\frac{(P(\bm y|{\bm{c}}))^{s}}{(P(\bm y|\bm 0))^{s}}\right]^\gamma\\
			&\overset{(*)}{=}2^{\ell\gamma}\left\{ \prod\limits_{i=0}^{n_\ell - w_\ell - 1}\sum\limits_{ y_i\in \mathcal{Y}}P(y_i|0)^{1-s\gamma}\Big[\sum\limits_{{c}_i\in \mathbb{F}_2} P({c}_i)(P(y_i|{c}_i))^{s}\Big]^\gamma\right\}
\left\{ \prod\limits_{i=n_\ell - w_\ell}^{n_\ell - 1}\sum\limits_{ y_i\in \mathcal{Y}}P(y_i|0)^{1-s\gamma}\Big[ P(1)(P(y_i|1))^{s}\Big]^\gamma\right\}\\
			&\overset{(**)}{=}\exp\left[- n_\ell \left( \frac{n_\ell - w_\ell}{n_\ell}E_0\left(\frac{1}{2},\gamma \right) + \frac{w_\ell}{n_\ell}E_0\left(1,\gamma\right) - \gamma R_\ell \right)   \right] \\
&\overset{(***)}{\leq}\exp(- n_\ell E(w_\ell, R_\ell)) ,
		\end{aligned}
	\label{Eq_main}
	\end{equation}
where the equality $(*)$ follows from the memoryless channel assumption and the distribution of the codewords in $\mathscr{C}_\ell$, the equality $(**)$ follows by recalling that $s=1/(1+\gamma)$ and the definition
	\begin{equation}
		\begin{aligned}
			E_0(p,\gamma)=-\log\Bigg\{\sum\limits_{ y\in \mathcal{Y}}P( y| 0)^{\frac{1}{1+\gamma}}&\Big[(1-p)(P( y| 0))^{\frac{1}{1+\gamma}}+p(P( y| 1))^{\frac{1}{1+\gamma}}\Big]^\gamma\Bigg\} ,
		\end{aligned}	
	\end{equation}
and the inequality $(***)$ follows by denoting
	\begin{equation}
		E(w_\ell, R_\ell)=\max\limits_{0\leq\gamma\leq 1}\left( \frac{n_\ell - w_\ell}{n_\ell}E_0\left(\frac{1}{2},\gamma \right) + \frac{w_\ell}{n_\ell}E_0\left(1,\gamma\right) - \gamma R_\ell \right)\text{.}
\label{Eq:E_ell}	
\end{equation}
Thus, we have
	\begin{equation}
			{\rm Pr}\{{\text{error}}|\bm 0\} \leq \sum_{\ell = 0}^{k-1}\exp\left[- n_\ell E(w_\ell,R_\ell) \right].
    \end{equation}
Since ${\rm Pr}\{{ \text{error}}|\bm u^k\} = {\rm Pr}\{{ \text{error}}|\bm 0\}$ for linear codes over BIOS channels, we have
\begin{equation}\label{Eq:FER}
  {\rm FER}_{\text{avg}} = \sum_{\bm u^k \in \mathbb{F}_2^{k} } 2^{-k} {\rm Pr}\{{ \rm error}|\bm u^k\} \leq  \sum_{\ell = 0}^{k-1}\exp\left[- n_\ell E(w_\ell,R_\ell) \right].
\end{equation}

\end{Proof}

\begin{theorem}
Let $k<n$ be two positive integers.
For two arbitrarily small positive numbers $\epsilon$ and $\delta$, one can always find a sufficiently large integer ${n}_T$ such that, for all $n \geq {n}_T$ and $k=\lceil   n(C_{\text{BIOS}} - \delta) \rceil$, there exists a staircase generator matrix $\textbf{G}$ satisfying that $C_{\text{BIOS}} - \delta \leq k/n < C_{\text{BIOS}}$ and the ML decoding error rate ${\rm FER} \leq \epsilon$.

\end{theorem}

\begin{Proof}
Let $0 < \alpha \leq 1$.
Set $w_0 = \lceil \alpha(n-(k-1)) \rceil$ and $w_i$ for $i =
0, 1, \cdots , k - 1$ such that $\sum_{i = 0}^{k-1} w_i = n$ and $w_0 \geq w_1 \geq \cdots \geq w_{k-1}\geq 1$.
Consider the SGMC ensemble with the profile $\bm{w} = (w_0,w_1,\cdots,w_{k-1})$.
For \eqref{Eq:E_ell}, we have $\frac{n_\ell - w_\ell}{n_\ell}E_0\left(\frac{1}{2}, 0 \right) + \frac{w_\ell}{n_\ell}E_0\left(1, 0 \right) - 0\cdot R_\ell=0$ and
	\begin{equation}
\begin{split}
   \frac{\partial \left(\frac{n_\ell - w_\ell}{n_\ell}E_0\left(\frac{1}{2},\gamma \right) + \frac{w_\ell}{n_\ell}E_0\left(1,\gamma\right)\right)}{\partial \gamma}-R_\ell \Bigg|_{\gamma=0} & =\frac{n_\ell - w_\ell}{n_\ell}I_{0}(\frac{1}{2}) + \frac{w_\ell}{n_\ell}I_0(1)-R_\ell, \\
     &\geq I_0(\frac{1}{2}) - R_\ell ,
\end{split}
	\end{equation}
where the inequality follows from that the partial mutual information $I_0(p)$ increases from $I_0(0)=0$ to $I_0(1)$~\cite{Ma2022ISIT}.
Since $w_0 \geq w_1 \geq \cdots \geq w_{k-1} \geq 1$, for $i < j$, we have
\begin{equation}\label{aaaa}
\begin{aligned}
   & j(w_0+w_1+\cdots+w_i) - i(w_0+w_1+\cdots+w_j) \\
 =  &  (j-i)(w_0+w_1+\cdots+w_i) - i(w_{i+1}+\cdots+w_j) \\
 >  &  (j - i)(i+1)w_i - (i+1)(j-i)w_{i+1} \\
 = & (j - i)(i+1)(w_i - w_{i+1}) \\
 \geq & 0
\end{aligned}.
\end{equation}
Hence, we have
\begin{equation}\label{aaaa}
 R_i = \frac{i}{w_0+w_1+\cdots+w_i} < \frac{j}{w_0+w_1+\cdots+w_j} = R_j,
\end{equation}
and hence $R_{0} < R_1 < \cdots < R_{k-1} = (k-1)/n < k/n < C_{\rm BIOS}$.
Hence, we have $E(w_\ell, R_\ell) > 0$ for all $\ell \geq 0$.
Now letting $n \rightarrow \infty$, we have $n_{\ell} \rightarrow \infty $ since $n_{\ell} \geq w_0 \geq  \alpha n(1- R)$ for $\ell \geq 0$.
Thus, every term of the summation $\sum_{\ell = 0}^{k-1}\exp\left[- n_\ell E(w_\ell,R_\ell)\right]$ of~\eqref{conditional_pro} in Theorem~\ref{theorem_1} goes to $0$ exponentially as $n_{\ell} \rightarrow \infty $ and the number of terms is $k = nR$~(polynomially increasing).
Hence, ${\rm FER}_{\text{avg}} \leq \sum_{\ell = 0}^{k-1}\exp\left[- n_\ell E(w_\ell,R_\ell)\right] \leq  \epsilon$ for sufficiently large $n$, implying that there exists at least a staircase generator matrix $\mathbf{G}$ of size $k \times n$ such that ${\rm  FER }  \leq   \epsilon$.
\end{Proof}

\section{Decoding Algorithm and Performance Analysis}\label{sec3}

\subsection{LC-ROSD}
The original OSD~\cite{fossorier1995soft} suffers from two main issues.
The first issue is the use of the GE, which is inevitable in determining the MRB. Even worse, for a general matrix, only serial~(column-by-column) implementation of GE is available, leading to an additional decoding latency.
The second issue is the sub-optimality of the search order on the basis of the Hamming weights of the TEPs, without exploiting the soft information.
Actually, for the seconde issue, with an optimized search order, several tailored early stopping criteria can be used to reduce the number of searches.
The second issue has been resolved by the LC-OSD~\cite{Wang2022LCOSD,Liang2023LCOSD} that selects an extended MRB of size $k+\delta$ and utilizes the SLVA~\cite{Seshadri1994} or the two-way FPT algorithm~\cite{liang2023randomarXiv}~(based on the FPT~\cite{wang2021semi}) by taking into the local constraints introduced by the extended MRB. The LC-OSD skips many invalid TEPs and hence reduces the number of searches down to less than ten from hundreds of thousands in the high SNR region.
The first issue can be mitigated for the SGMC by the use of representative OSD with local constraints~(LC-ROSD)~\cite{wang2023SGMC}, which selects at least one bit from each staircase to form a locally reliable basis, resulting in a lower triangular sub-matrix with ones on the diagonal.
This lower triangular sub-matrix enables parallel implementation of the GE process, supporting a potentially low decoding latency.
The LC-ROSD algorithm~\cite{wang2023SGMC} is re-described below for completeness.

Upon receiving the vector $\bm{y}$, the log-likelihood ratio~(LLR)~sequence
$\bm{\lambda}$ can be calculated as
$$\lambda_i =
  \log\left(\frac{P(y_i |c_i = 0)}{P(y_i |c_i = 1)}\right), ~~~ 0 \leq i \leq n-1,
$$
and the hard-decision sequence $\bm{z}$ is given by
$$
    z_j =
\begin{cases}
1,  & \mbox{if }\lambda_j<0 \\
0, & \mbox{if }\lambda_j\geq0
\end{cases},~~~ 0 \leq i \leq n-1.
$$
The absolute value $|\lambda_i|$ is referred to as the reliability of $z_i$.
A test codeword $\bm{v} \in \mathbb{F}_2^n $ is given by $\bm{v}=\bm{z} + \bm{e}$, where $\bm{e} \in \mathbb{F}_2^n$ denotes the hypothetical test error pattern~(TEP) and ``+'' denotes the addition over $\mathbb{F}_2^n$.

\begin{enumerate}
  \item For a preset parameter $\delta$, find a permutation matrix $\bf \Pi$ such that the column-interleaved matrix
  \begin{equation}
    \label{equ:gauss-elimmu}
    \bf G\Pi=
    \begin{bNiceArray}{cw{c}{1.4cm}|cw{c}{1.4cm}|cw{c}{1.4cm}}[margin, first-row, last-col]
      \Block{1-2}{_{k\textrm{ columns}}} & & \Block{1-2}{_{\delta\textrm{ columns}}} & &\Block{1-2}{_{n-k-\delta\textrm{ columns}}}\\
      \Block{3-2}{\mathbf{L}} & &\Block{3-2}{\mathbf{Q}_1} & & \Block{3-2}{\mathbf{Q}_2} & &\Block{3-1}{^{\rotate k\textrm{ rows}}}\\
      & & & \\
      & & & \\
      \end{bNiceArray}
  \end{equation}
satisfies the following two constraints.
\begin{itemize}
  \item [i)] The $i$-th column of matrix $\mathbf{L}$ corresponds the most reliable bit within the $i$-th staircase. Precisely, it corresponds to the most reliable coordinate $t$ in the range of $\sum_{0\leq m<i}w_m\leq t <\sum_{0\leq m<(i+1)}w_m$. Notice that $\mathbf{L}$ is a full rank matrix since $\mathbf{L}$ is a lower triangular matrix with ones on the diagonal.
  \item [ii)] The $\delta$ columns of ${\bf Q}_1$ correspond to the most reliable $\delta$ bits among the remaining $n-k$ bits.
\end{itemize}
  \item Perform the parallel GE algorithm~(Algorithm \ref{algorithm:GE}), resulting in the form \begin{equation}
    \label{equ:gauss-elimG-SGMCs}
    \widetilde{\mathbf{G}}=
    \begin{bNiceArray}{cw{c}{1.4cm}|cw{c}{1.4cm}|cw{c}{1.4cm}}[margin, first-row, last-col]
      \Block{1-2}{_{k\textrm{ columns}}} & & \Block{1-2}{_{\delta\textrm{ columns}}} & &\Block{1-2}{_{n-k-\delta\textrm{ columns}}}\\
      \Block{3-2}{\mathbf{I}} & &\Block{3-2}{\mathbf{P}_1} & & \Block{3-2}{\mathbf{P}_2} & &\Block{3-1}{^{\rotate k\textrm{ rows}}}\\
      & & & \\
      & & & \\
      \end{bNiceArray}.
  \end{equation}
  \item  Permute the LLR vector $\bm{\lambda}$, the hard-decision sequence $\bm{z}$, the TEP $\bm{e}$ and the test candidate $\bm{v} = \bm{z} - \bm{e}$ into $\bm{\widetilde{\lambda}}$, $\bm{\widetilde{z}}$, $\bm{\widetilde{v}}$ and $\bm{\widetilde{e}}$ by the permutation matrix $\bm{\Pi}$.

  \item

 Rewrite $\bm{\widetilde{e}}=(\bm{\widetilde{e}}_L,\bm{\widetilde{e}}_M,\bm{\widetilde{e}}_R)$ with $\bm{\widetilde{e}}_L$ of length $k$, $\bm{\widetilde{e}}_M$ of length $\delta$ and $\bm{\widetilde{e}}_R$ of length $(n-k-\delta)$. Rewrite $\bm{\widetilde{z}}=(\bm{\widetilde{z}}_L,\bm{\widetilde{z}}_M,\bm{\widetilde{z}}_R)$ in the same way. Given the systematic  $\widetilde{\mathbf{G}}$, we have
 \begin{equation}\label{Eq:reencoding}
   (\widetilde{\bm{z}}_L + \widetilde{\bm{e}}_L) \widetilde{\mathbf{G}} = \widetilde{\bm{v}} = \widetilde{\bm{z}} + \widetilde{\bm{e}},
 \end{equation}
 and thus
      \begin{equation}
    \begin{cases}
      \bm{\widetilde{z}}_L{\bf P}_1+\bm{\widetilde{e}}_L{\bf P}_1= \bm{\widetilde{z}}_M+\bm{\widetilde{e}}_M\\
      \bm{\widetilde{z}}_L{\bf P}_2 +\bm{\widetilde{e}}_L{\bf P}_2=\bm{\widetilde{z}}_R+\bm{\widetilde{e}}_R
      \end{cases},
  \end{equation}
  indicating that $\bm{\widetilde{e}}_R$ and hence $\bm{\widetilde{e}}$ can be determined by $\bm{\widetilde{e}}_L$ and $\bm{\widetilde{e}}_M$.
  Hence, we can search $(\bm{\widetilde{e}}_L, \bm{\widetilde{e}}_M)$ instead of $\bm{\widetilde{e}}$.
  This can be done efficiently by performing the SLVA~\cite{Seshadri1994} over a trellis specified by the sub-matrix ${\bf P}_1$ or the two-way FPT algorithm~\cite{liang2023randomarXiv} with the local constraint $\bm{\widetilde{z}}_L{\bf P}_1+\bm{\widetilde{e}}_L{\bf P}_1= \bm{\widetilde{z}}_M+\bm{\widetilde{e}}_M$ in
  an order such that the partial soft weights $\Gamma(\bm{\widetilde{e}}_L)+\Gamma(\bm{\widetilde{e}}_M)$ are non-decreasing, where the soft weight of a TEP $\bm{e}=\bm{z}-\bm{v}$ is defined by
  \begin{equation}
    \label{equ:soft-weight}
    \begin{split}
      \Gamma(\bm{e})
      = \log\frac{\Pr\{\bm{y}\mid\bm{c}=\bm{z}\}}{\Pr\{\bm{y}\mid\bm{c}=\bm{v}\}}
      &= \sum_{i=0}^{n-1} \log{\frac{\Pr\{y_i\mid c_i=z_i\}}{\Pr\{y_i\mid c_i=z_i-e_i\}}} \\
      &= \sum_{i=0}^{n-1} e_i|\lambda_i|
      \triangleq \langle\bm{e},|\bm{\lambda}|\rangle
    \end{split}.
  \end{equation}
\end{enumerate}

\begin{algorithm}[tb]
  \renewcommand{\algorithmicrequire}{\textbf{Input:}}
  \renewcommand{\algorithmicensure}{\textbf{Output:}}
  \caption{Parallel Gaussian Elimination}
  \label{algorithm:GE}
  \begin{algorithmic}[1] 
  \Require ~~\\ 
       A matrix $\mathbf{G} = \begin{bmatrix} \mathbf{L} & \mathbf{Q} \end{bmatrix}$ of size $k \times n$, where $\mathbf{L}$ is a lower triangular matrix with ones on the diagonal.
      \Ensure ~~\\ 
      A matrix $\bf \widetilde{G}=\begin{bmatrix}{\bf I}& {\bf P} \end{bmatrix}$ of size $k\times n$.

      \State Set ${\bf \widetilde{G}}(0,:)  \leftarrow {\bf G}(0,:)$. \Comment{Denote by ${\bf G}(\ell,:)$ the $\ell$-th row of ${\bf G}$.}
      \State For $\ell \geq 1$, perform the following loop, denoted as RowReduction~$(\ell)$, in parallel.
      ~~\Comment{
      Denote by ${\bf \widetilde{G}}(\ell,i)$ the $i$-th element of ${\bf \widetilde{G}}(\ell,:)$.}
      \Function{RowReduction}{$\ell$}
      \State Initialize ${\bf \widetilde{G}}(\ell,:)  \leftarrow {\bf G}(\ell,:)$.
      \For{$i= \ell-1,\ell-2,\cdots,0$}
        \If{${\bf \widetilde{G}}(\ell,i)=1$}
          \State ${\bf \widetilde{G}}(\ell,:) \leftarrow {\bf \widetilde{G}}(\ell,:)+{\bf G}(\ell,:)$.
        \EndIf
      \EndFor
\EndFunction
\Return $\bf \widetilde{G}$.

  \end{algorithmic}
\end{algorithm}

\begin{figure}[tp]
  \centering
  \subfloat[FER\label{fig:compare-ohterOSD-fer}]{\includegraphics[width=\figwidth]{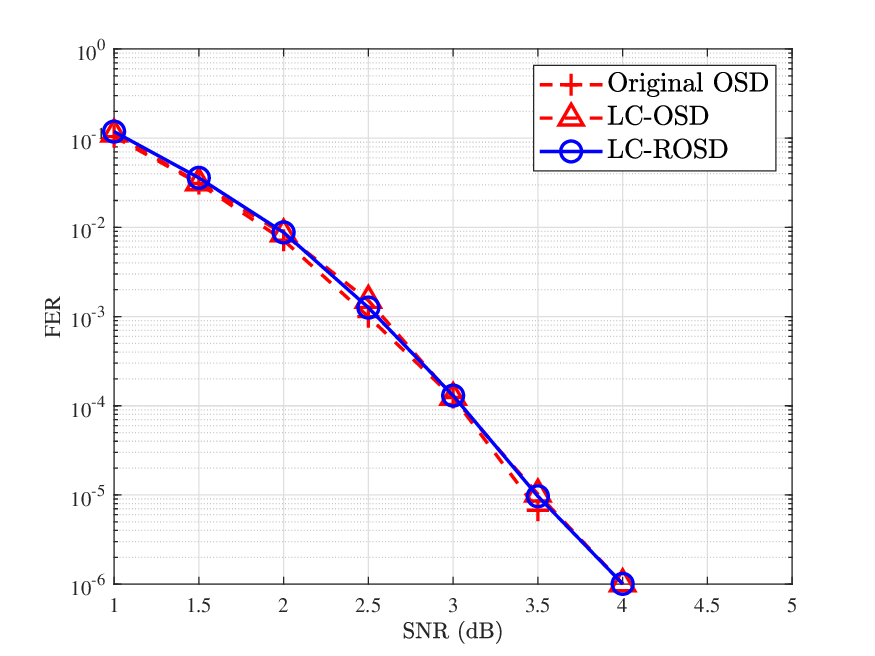}}
  \\
  \subfloat[Average number of TEPs\label{fig:compare-ohterOSD-tep}]{\includegraphics[width=\figwidth]{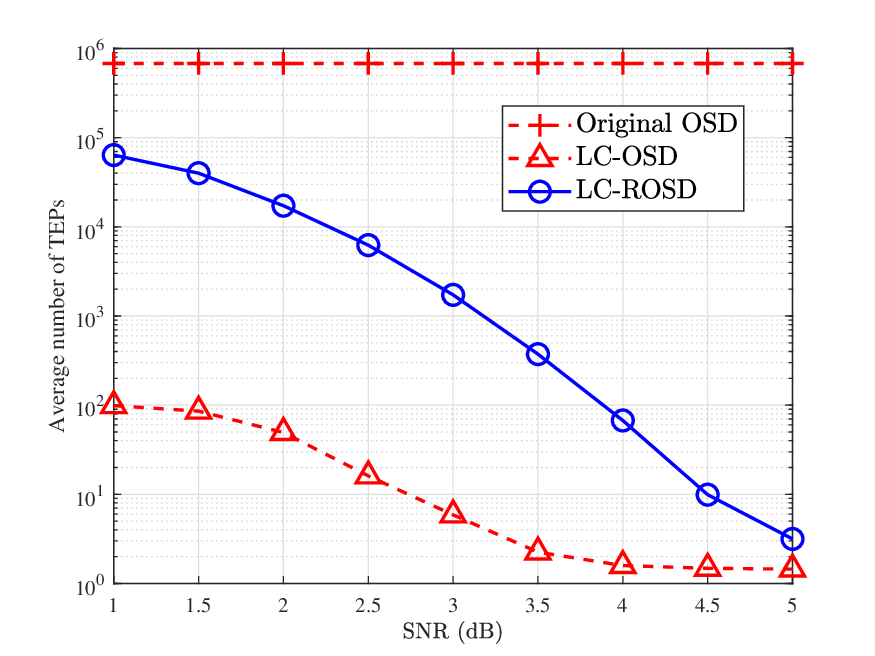}}
  \caption{
    Performance comparison of the proposed SGMC~$\mathscr{C}[128,64]$ under the original OSD~(order $t = 4$), the LC-OSD~($\delta = 12$ and $\ell_{\max} = 10^6$) and the LC-ROSD~($\delta = 12$ and $\ell_{\max} = 10^6$).
  }
  \label{fig:Compare-other-OSD}
\end{figure}

\begin{example}
Consider an SGMC $\mathscr{C}[128, 64]$~(randomly generated but fixed) with the same profile as the RM code $[128,64]$, where the codewords are transmitted with the BPSK modulation over the AWGN channels.
We have simulated three decoding algorithms: the original OSD~\cite{fossorier1995soft}, the LC-OSD~\cite{Liang2023LCOSD} and the LC-ROSD.
The simulation results are shown in Fig.~\ref{fig:Compare-other-OSD}, from which we
see that the LC-ROSD has a similar performance to the original OSD but has a much smaller number of TEPs than the original OSD.
The slight increase in the average number of TEPs for the LC-ROSD over the LC-OSD can be justified by the fact that the basis for LC-ROSD is locally reliable and the basis for LC-OSD is globally reliable.
Actually, the slight increase in the number of TEPs is the cost caused by the use of the low complexity~(parallel) GE, which provides a way for complexity trade-off.
The LC-ROSD can be used in the high SNR region~(where GE is the main cause of latency), while the LC-OSD can be used in the low-to-moderate SNR region~(where re-encoding for numerous TEPs is the main cause of latency).
\end{example}

\subsection{Weight Spectrum for the Random Staircase Generator Matrix Codes}\label{SecIIIB}
Recalling that the subset $\mathcal{U}_\ell \subset \mathbb{F}_2^k$ consists of $2^\ell$ sequences whose right-most 1 appears at the position $\ell$, we can derive the partial weight spectrum for $\bm{c}\in \mathscr{C}_\ell = \{\bm c= \bm u {\bf G}|{\bm u} \in \mathcal{U}_\ell\}$ average over the ensemble as
\begin{equation}\label{Eq:weight-spectrum}
  B^{(\ell)}(X) =  2^\ell X^{w_\ell}(1/2+1/2X)^{n_\ell - w_\ell},
\end{equation}
as tabulated in Table~\ref{Table:Weight_Spectrum}.
This is because, once $\bm{h}_\ell$ is totally random, any
partial codeword (the left part of length $n_\ell - w_\ell$) must be totally random no matter what other elements are.
Hence, the complete ensemble weight spectrum is given by
\begin{equation}\label{Eq:weight_spectrum}
B(X) = 1 + \sum_{\ell = 0}^{k-1}B^{(\ell)}(X) = 1 + \sum_{\ell = 0}^{k-1} 2^\ell X^{w_\ell}(1/2 + 1/2X)^{n_\ell - w_\ell}.
\end{equation}

\begin{figure}[tp]
    \centering
    \includegraphics[width=\figwidth]{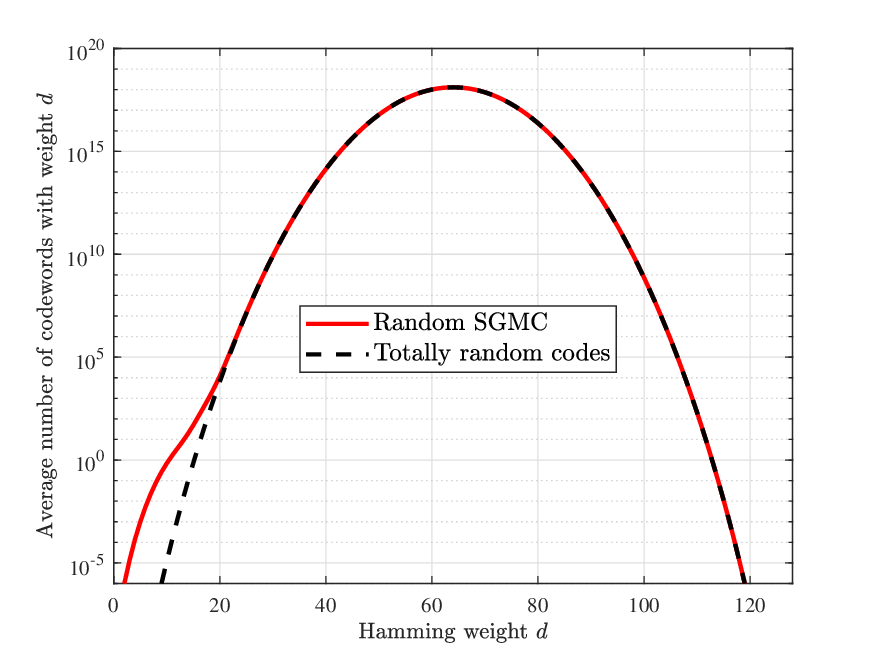}
    \caption{The ensemble weight spectrum of the proposed random SGMC $[128, 64]$ with the RM profile versus that of the totally random code $[128, 64]$.}
    \label{fig:ImprovedNU-weight}
\end{figure}

\begin{example}
The ensemble weight spectrum of the proposed random SGMC~$\mathscr{C}[128,64]$ with the RM profile is shown in Fig.~\ref{fig:ImprovedNU-weight}. Also plotted in Fig.~\ref{fig:ImprovedNU-weight} is the ensemble weight spectrum of the totally random code\footnote{By a totally random code, we mean in this paper a linear code with a generator matrix whose elements are all sampled from the Bernoulli process with success probability $1/2$.}~of length $n=128$ and dimension $k=64$, given by
\begin{equation}\label{Eq:}
 A(X) =  1 + (2^k-1)(1/2+1/2X)^n.
\end{equation}
From Fig.~\ref{fig:ImprovedNU-weight} we observe that the ensemble weight spectrum of the random SGMC matches well with that of the totally random codes except in the low-weight region.
This deviation is due to the semi-randomness of the lower triangular generator matrix.
\end{example}

\begin{table}[tp]
    \centering
    \caption{Partial weight spectrum for different input.}
    \begin{tabular}{c|c}
    \hline
        input & average partial weight spectrum for $\mathscr{C}_\ell$  \\ \hline \hline
        $\bm{u} = \bm{0}$ & 1 \\ \hline
        $\bm{u} \in \mathcal{U}_{0}$ & $ X^{w_0}$ \\ \hline
        $\bm{u} \in\mathcal{U}_{1}$ & $ 2 X^{w_1} (1/2 + 1/2X)^{w_0}$ \\  \hline
        $\bm{u} \in\mathcal{U}_{2}$ & $ 2^2 X^{w_2} (1/2 + 1/2X)^{w_0 +w_1}$ \\ \hline
                $\vdots$ & \vdots \\ \hline
    $\bm{u} \in\mathcal{U}_{k-1}$ & $ 2^{k-1} X^{w_{k-1}} (1/2 + 1/2X)^{\sum_{j = 0}^{k-2} w_j}$ \\  \hline
    \end{tabular}
    \label{Table:Weight_Spectrum}
\end{table}

The minimum Hamming weight of the SGMC, denoted by $d_{\min}$, is crucial for the performance in the high SNR region.
It can be upper bounded by the minimum Hamming weight $d_{\min, j}$ of $\mathcal{D}_j = \bigcup_{\ell<j}\mathscr{C}_\ell$ for $j\geq 1$.
Here, $\mathcal{D}_j \bigcup \{\bm 0\}$ is the $j$-dimensional sub-code expanded the first $j$ rows of the generator matrix.
Thus, we have $d_{\min} = d_{\min,k} \leq  d_{\min,k-1} \leq \cdots \leq d_{\min,1} = w_0$.
Although the upper bounds become tighter as $j$ increases, we consider $d_{\min,2}$ of $\mathcal{D}_2$ with $w_0 \geq w_1$ for simplicity\footnote{Intuitively, the codewords with the minimum Hamming weight occurs in $\mathcal{D}_2$ with a high probability.}.
The minimum Hamming weight $d_{\min,2}$, as a random variable over the code ensemble, takes values between $w_1$ and $w_1 + \min(\lfloor w_0/2 \rfloor,  w_0- w_1)$ inclusive with the probability mass function given by
\begin{equation}\label{Eq:dmin2-i0}
  \Pr\{d_{\min,2} = w_1 + i\} = \binom{w_0}{i}2^{-w_0+1} ~~~~~~~~~~~~~{\rm for}~ i < \min(\lfloor w_0/2 \rfloor,  w_0- w_1),
\end{equation}
and
\begin{equation}\label{Eq:dmin2-i1}
  \Pr\{d_{\min,2} = w_1 + i\} = 1 - \sum_{j < i}\binom{w_0}{j}2^{-w_0+1} ~~~{\rm for}~ i = \min(\lfloor w_0/2 \rfloor,  w_0- w_1).
\end{equation}

\begin{figure}[tp]
    \centering
    \includegraphics[width=\figwidth]{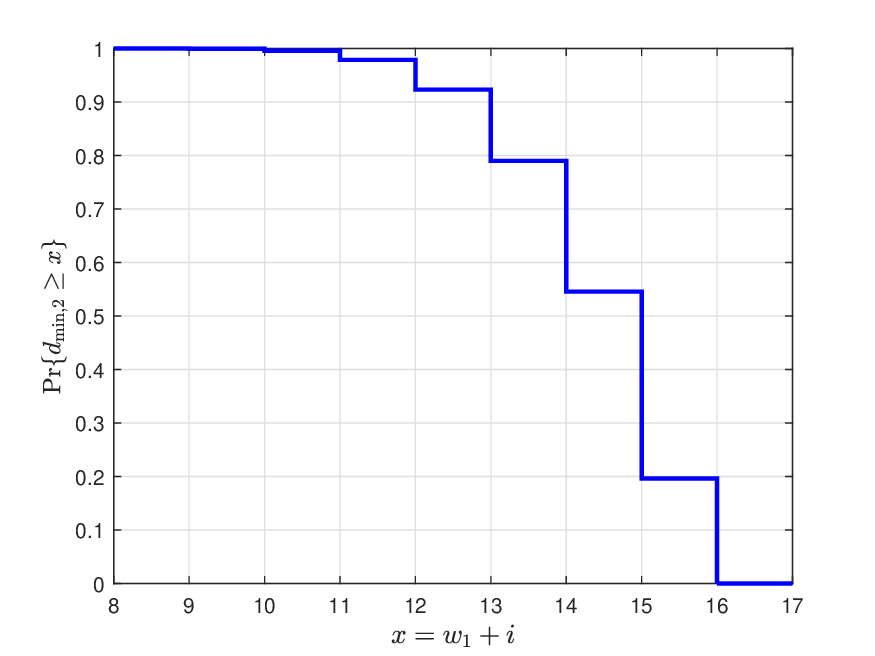}
    \caption{For  $w_0 = 16$ and $w_1$ = 8, the complementary cumulative distribution function $\Pr\{d_{\dim,2} \geq w_1 +i\}$ for $0 \leq i \leq 8$.}
    \label{fig:dmin}
\end{figure}

\emph{{Example \rm{3}~(Cont’d)}}:  Consider the RM profile with $w_0 = 16$ and $w_1 = 8$. The complementary cumulative distribution function $\Pr\{d_{\min,2} \geq w_1 + i\}$ for $0 \leq i \leq 8$ is shown in Fig.~\ref{fig:dmin}, from which we observe that the range of $d_{\min,2}$ is between 8 and 16, and that over $90\%$ of randomly generated SGMCs have $d_{\min, 2} \geq 13$.

The ensemble weight spectrum can be used to derive the union bound~(UB) on the FER~\cite{ma2013UB}, as illustrated in the following example.

\begin{figure}[tp]
    \centering
    \includegraphics[width=\figwidth]{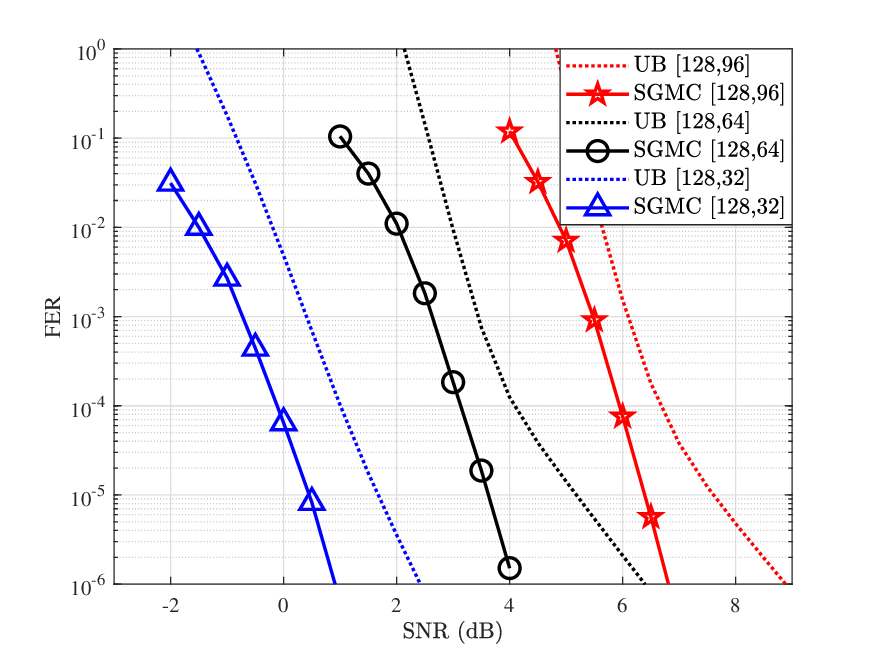}
    \caption{Performance of the proposed SGMCs with the RM-profiles of length $n =128$ and $k \in \{32,64,96\}$, where the LC-ROSD with $\delta = 12$ and $\ell_{\max} =10^6$ is employed for decoding.}
    \label{fig:UB-performance}
\end{figure}

\begin{example}
Consider SGMCs~(randomly generated but fixed) with the RM-profiles of length $n = 128$ and dimension $k \in \{32,64,96\}$.
The codewords are transmitted with the BPSK modulation over the AWGN channels and the LC-ROSD with $\ell_{\max} = 10^6$ is employed for the decoding.
The simulation results are shown in Fig.~\ref{fig:UB-performance}, from which we observe that the UB is loose, especially in the high SNR region.

\end{example}

\subsection{Partially Random Coding Union Bound}
Similar to but different from the RCU bound~\cite{polyanskiy2010channel}, which applies to totally random codes, we derive a partially RCU bound, which applies to semi-random codes like SGMCs.

Without loss of generality, suppose that the zero codeword $\bm 0$ is transmitted over a BIOS channel, resulting in $\bm Y$, where elements under the staircase of ${\bf G}$ are sampled from a Bernoulli process with success probability $1/2$.
Given the partition of the message, denote by ${\bm X}_\ell = ({\bm X}_{\ell,a},{\bm 1}^{w_\ell},{\bm 0}^{n - n_\ell}) = {\bm U}_\ell{\bf G}\in \mathscr{C}_\ell$ for ${\bm U}_\ell \in \mathcal{U}_\ell$, where ${\bm X}_{\ell,a}$ of length $\sum_{i< \ell}w_i$ is sampled from a Bernoulli process with success probability $1/2$,
${\bm 1}^{w_\ell}$ is the all-one vector of length $w_\ell$ and ${\bm 0}^{n - n_\ell}$ is the all-zero vector of length $n - n_\ell = n -  \sum_{i\leq \ell}w_i$.
Similar to~\cite[Theorem~16]{polyanskiy2010channel}, the probability that the ML decoding output is in $\mathscr{C}_\ell~(\ell \geq 0)$ can be upper bounded by
\begin{equation}\label{Eq:RCU}
{\rm RCU}_\ell[n_\ell,R_\ell]\! \triangleq \! \mathbb{E} \left[ \min\{1,\!(2^{n_\ell R_\ell}\!\!-1){\rm PEP}_\ell(\bm{Y}) \} \right],
\end{equation}
where the expectation is taking over the received vector $\bm{Y}$ given that the all zero codeword is transmitted and ${\rm PEP}_\ell(\bm{Y})$ is the pairwise error probability conditional on $\bm{Y}$, measured over the defined SGMC ensemble as
\begin{equation}\label{Eq:PEP}
  \begin{aligned}
  & {\rm PEP}_\ell(\bm{Y})  \triangleq \Pr \{ P(\bm{Y}|{\bm{X}}_\ell) \geq P(\bm{Y}|{\bm{0}})|\bm Y \} \\
  & = \Pr \left\{ P(\bm{Y}_a|{\bm{X}}_{\ell,a}) \geq \frac{P(\bm{Y}_a,\bm{Y}_b|{\bm{0}}^{n_\ell})}{ P(\bm{Y}_b|{\bm{1}}^{w_\ell})} |\bm{Y}_a,\bm{Y}_b \right\}
  \end{aligned},
\end{equation}
which can be approximated by the saddlepoint approximations~\cite{font2018saddlepoint}, where $\bm{Y} = (\bm{Y}_a, \bm{Y}_b, \bm{Y}_c)$ is rewritten in the same way as ${\bm{X}}_\ell$.
Thus, we have\begin{equation}\label{Eq:UpperBound}
   {\rm FER}_{\rm avg}   \leq \sum_{\ell = 0}^{k - 1} {\rm RCU}_{\ell}[n_\ell,R_\ell],
\end{equation}
which is referred to as the partially RCU bound. Notice that ${\rm RCU}_0[w_0, R_0]$ can be replaced~(hence tightened) by its actual value, i.e., the FER of the repetition code of length $w_0$.

\begin{figure}[tp]
    \centering
    \includegraphics[width=\figwidth]{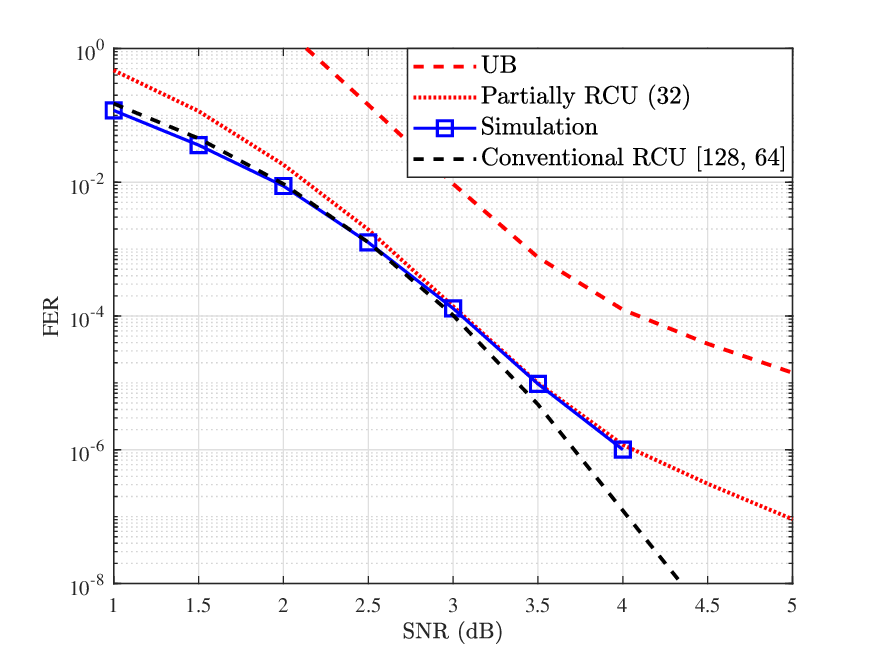}
    \caption{Performance of the proposed SGMC~$[128,64]$ with the RM-profile, where LC-ROSD with $\delta = 12$ and $\ell_{\max} = 10^6$ is employed for decoding.}
    \label{fig:RM-FER-UB}
\end{figure}

\begin{example}
Consider an SGMC $\mathscr{C}[128, 64]$~(randomly generated but fixed) with the same profile as the RM code $[128,64]$.
The codewords are transmitted with the BPSK modulation over the AWGN channels and the LC-ROSD  is employed for the decoding.
The simulation results are shown in Fig.~\ref{fig:RM-FER-UB}, where ${\rm RCU}_0[n_0, R_0]$ has been tightened by its actual  value  $Q(w_0/\sigma)$. We have the following observations.
In the low-to-moderate SNR region, the decoding performance can approach the conventional RCU bound~\cite{polyanskiy2010channel} for the totally random code $\mathscr{C}[128, 64]$. As the SNR increases, the decoding performance deviates from the conventional RCU bound but matches well with the proposed partially RCU bound~\eqref{Eq:UpperBound}, which is tighter than the conventional UB. The gap away from the conventional RCU bound is about $0.3$~dB at the ${\rm FER} = 10^{-6}$.
This deviation, again, can be attributed to the semi-randomness of the considered SGMC.

\end{example}

\subsection{Lower Bounds}
In this subsection, we turn to the second-order Bonferroni lower bound~\cite{Cohen2004,Zheng2023LB} for the random SGMCs.
Suppose that the all-zero codeword $\bm{c}_{0} = \bm{0}$ is transmitted.
Given a received sequence $\bm y$,
let $E_i$ be the event that $P(\bm y|{\bm c}_i)\geq P(\bm y|\bm 0)$, where $\bm{c}_i ~(1\leq i \leq 2^{k-1} )$ denotes the $i$-th codeword.
Thus, the ML decoding error probability is denoted as
\begin{equation}\label{Eq:MLError}
   {\Pr}\{{\text{ML~decoding~is~error}}\} =\Pr \left\{ \bigcup_{i = 1}^{2^k - 1} E_i \right\}.
\end{equation}
For an arbitrary subset $\mathcal{J} \subseteq \{1,2,\cdots,2^k-1 \}$, we have
\begin{equation}\label{Eq:}
  \Pr \left\{ \bigcup_{i = 1}^{2^k - 1} E_i \right\} \geq \Pr \left\{ \bigcup_{j \in \mathcal{J}} E_j \right\}.
\end{equation}
From the second-order Bonferroni inequality~\cite{hoppe1985iterating}, we have
\begin{equation}\label{Eq:Bonferroni}
  \Pr \left\{ \bigcup_{j \in \mathcal{J}} E_j \right\} \geq \sum_{i \in \mathcal{J}} \Pr\left\{ E_i \right\} - \sum_{i<j (i,j \in \mathcal{J})} \Pr\left\{ E_i \bigcap E_j \right\},
\end{equation}
where $\Pr\{ E_i \}$  and $\Pr\{ E_i \bigcap E_j \}$  can be computed by~\cite{sason2006performance}
\begin{equation}\label{Eq:A_i}
  \Pr\{ E_i \} = \frac{1}{\pi}\int_{0}^{\pi/2} \exp \left[-\frac{W_H(\bm{c}_i)}{\sigma^2 \sin^2 \theta}\right]d \theta ,
\end{equation}
and
\begin{equation}\label{Eq:A_iA_j}
  \Pr\left\{ E_i \bigcap E_j \right\} = \psi \left( \frac{\sqrt{W_H(\bm{c}_i)}}{\sigma}, \frac{\sqrt{W_H(\bm{c}_j)}}{\sigma} \right),
\end{equation}
where
\begin{equation}\label{Eq:Psi}
  \begin{aligned}
  \psi(x,y) = &\frac{1}{2\pi}\int_0^{\pi/2 - \tan^{-1}(y/x)} \frac{\sqrt{1-\rho_{ij}}}{1-\rho_{ij}\sin2\theta}\exp\left[-\frac{x^2}{2} \frac{1-\rho_{ij}\sin2\theta}{(1-\rho_{ij}^2)\sin^2 \theta}\right]  d  \theta  \\ &+ \frac{1}{2\pi}\int_0^{\tan^{-1}(y/x)} \frac{\sqrt{1-\rho_{ij}}}{1-\rho_{ij}\sin2\theta}\exp\left[-\frac{y^2}{2} \frac{1-\rho_{ij}\sin2\theta}{(1-\rho_{ij}^2)\sin^2 \theta} \right]  d  \theta ,
\end{aligned}
\end{equation}
\begin{equation}\label{Eq:rhoij}
  \rho_{ij}= \frac{W_H(\bm{c}_i) + W_H(\bm{c}_j) - W_H(\bm{c}_i - \bm{c}_j)}{2\sqrt{W_H(\bm{c}_i)  W_H(\bm{c}_j)}},
\end{equation}
and $W_H(\cdot)$ denotes the Hamming weight function.

\begin{figure}[tp]
    \centering
    \includegraphics[width=\figwidth]{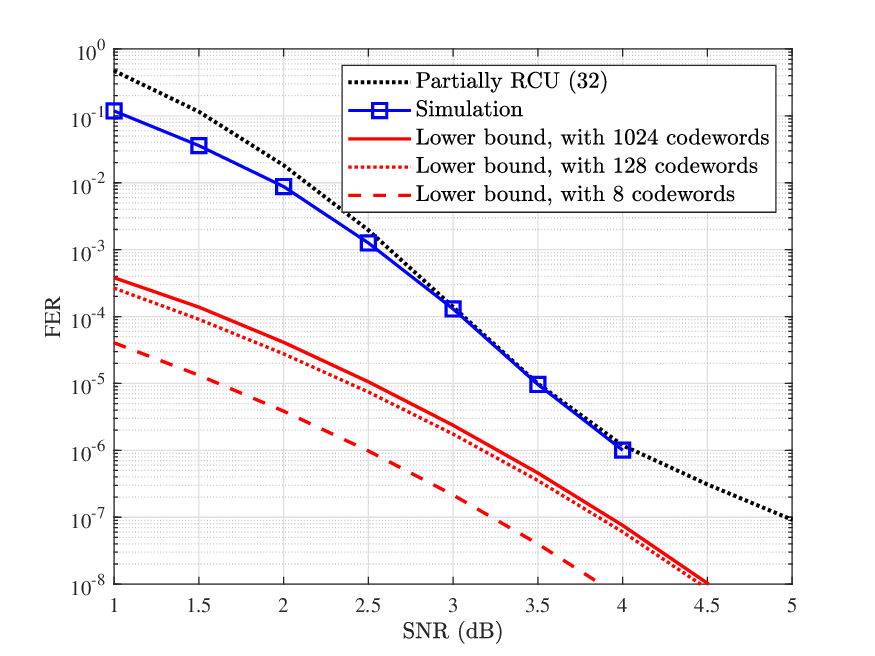}
    \caption{Performance of the proposed SGMC $[128,64]$ with the RM profile, where the LC-ROSD with $\delta = 12$ and $\ell_{\max} = 10^6$ is employed for decoding.}
    \label{fig:LowerBound-FER-ell}
\end{figure}

\begin{figure}[tp]
    \centering
    \includegraphics[width=\figwidth]{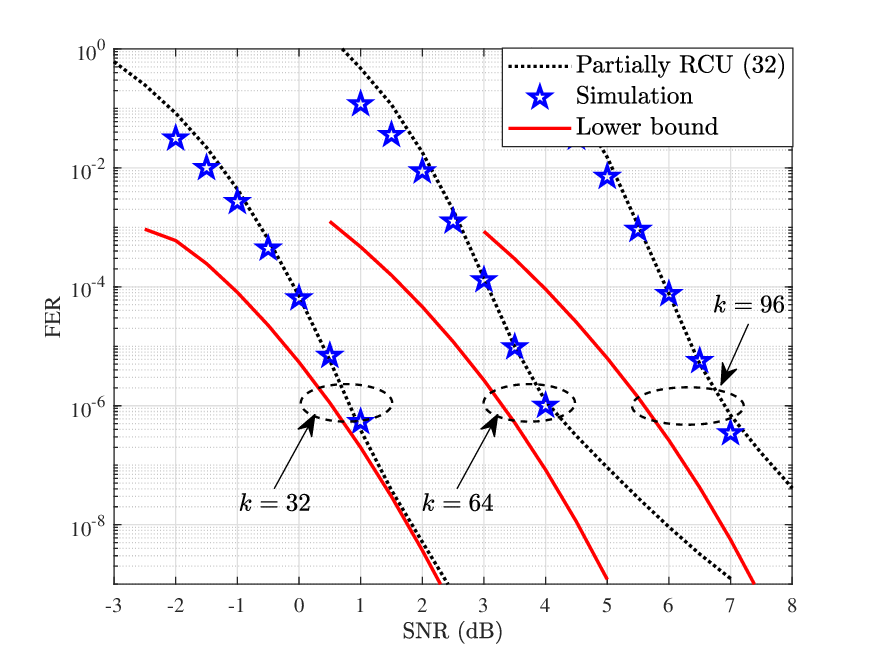}
    \caption{Performance of the proposed SGMCs with the RM-profiles of length $n =128$ and $k \in \{32,64,96\}$, where the LC-ROSD is employed for decoding.}
    \label{fig:LowerBound-FER}
\end{figure}

\begin{example}
Consider SGMCs~(randomly generated but fixed) with the RM-profiles, where the code length $n = 128$ and the dimension $k \in \{32,64,96\}$.
The codewords are transmitted with the BPSK modulation over the AWGN channels and the LC-ROSD is employed for the decoding.
The lower bounds with different number of codewords are shown in Fig.~\ref{fig:LowerBound-FER-ell}, where the involved codewords for computation are selected from low dimensional sub-codes. We observe that the lower bound becomes tighter as the number of involved codewords increases.
The numerical results for different code rates are shown in Fig.~\ref{fig:LowerBound-FER},
from which we observe that the decoding performance for different code rates can match well with the corresponding partially RCU bounds, approaching the corresponding lower bounds~(all with 1024 codewords) in the high SNR region. Hence, these tight lower bounds can be taken as benchmarks to demonstrate the near optimality of the decoding algorithm.

\end{example}

\section{Code Design and Performance Comparison}\label{sec4}

\subsection{Code Design}\label{Subsetion-III-E}
The performance is closely related to the profile, as evidenced by the ensemble weight spectrum and the partially RCU bound.
On the one hand, for the LC-ROSD of SGMCs, the relatively reliable basis is formed by selecting representative bits from each staircase. Hence, heuristically, we may widen the narrowest staircase to improve the quality of the basis.
This can be achieved by setting all staircases as uniformly as possible.
On the other hand, the minimum Hamming weight, as a key factor of the performance, is upper bounded by the row weights of the generator matrix.
So, $w_0$ should be large enough to guarantee the minimum Hamming weight.
Taking into account the aforementioned two aspects, we propose the following nearly uniform~(NU) profile.

\textbf{Definition 3~(Nearly Uniform Profile)}.
For any given $w_0 \geq \lceil {n}/{k} \rceil$, a profile $\bm w = (w_0, w_1, \cdots, w_{k-1})$ is referred to as the nearly uniform profile if $w_1 = w_2 = \cdots = w_j = \lceil (n-w_0)/(k-1) \rceil$, $w_{j+1} = \cdots = w_{k-1} = \lfloor (n-w_0)/(k-1) \rfloor$, and $j \lceil (n-w_0)/(k-1) \rceil + (k-1-j) \lfloor (n-w_0)/(k-1) \rfloor = n-w_0$.

\textbf{A code design recipe}: For a target FER and a tolerant gap, minimize $w_0$ such that the gap between the corresponding partially RCU bound and the conventional RCU bound is tolerant at the target FER.

\begin{figure}[tp]
    \centering
    \includegraphics[width=\figwidth]{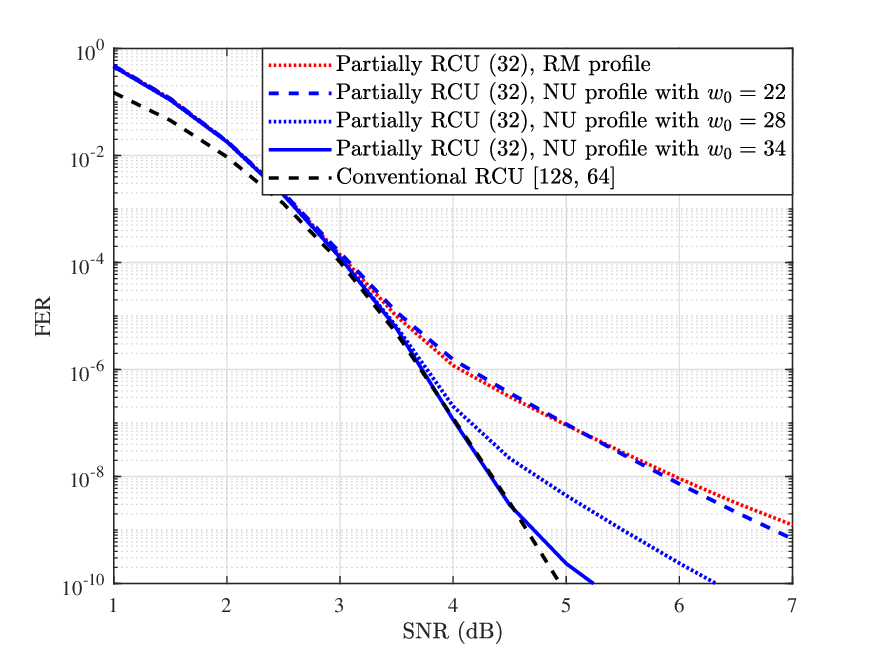}
    \caption{Partially RCU bound of the random SGMCs $[128,64]$ with the RM profile and the NU profiles.}
    \label{fig:RCU-RM-NU123}
\end{figure}

\begin{example}
  Consider SGMCs~$[128, 64]$. Shown in Fig.~\ref{fig:RCU-RM-NU123} are the partially RCU bounds corresponding to RM profile and different NU profiles, which can be used to guide the choice of $w_0$. For example, at the target FER of $10^{-5}$, $10^{-7}$ and $10^{-9}$, we may choose, respectively, the NU profiles with $w_0$ of $22$, $28$ and $34$.
\end{example}

\begin{figure}[tp]
    \centering
    \includegraphics[width=\figwidth]{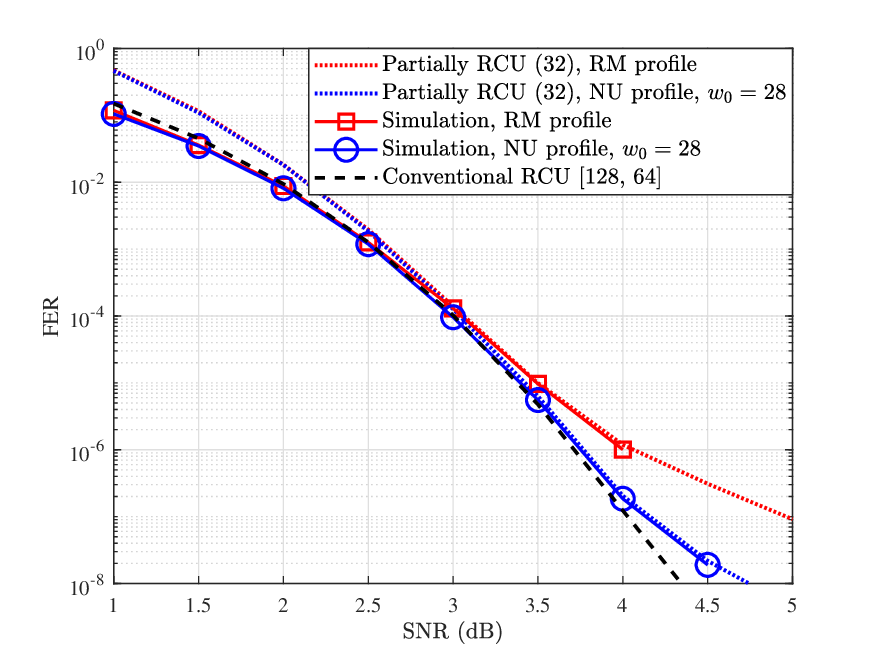}
    \caption{Performance of the proposed SGMCs $[128,64]$ with the RM profile and the NU profile, where the LC-ROSD with $\delta = 12$ and $\ell_{\max} = 10^6$ is employed for decoding.}
    \label{fig:ImprovedNU-FER}
\end{figure}

\begin{example}
Consider the SGMCs~$\mathscr{C}[128,64]$~(randomly generated but fixed) with the RM profile and the proposed NU profile, where the codewords are transmitted with BPSK modulation over the AWGN channels.
The RM profile is $\bm{w} = (16,8,4,2,1,1,8,4,\cdots)$, while the NU profile is $\bm{w} = (28,2,2,\cdots,2,1,1,\cdots,1)$.
The simulation results are shown in Fig.~\ref{fig:ImprovedNU-FER}, from which we observe that the decoding performance matches well with the proposed partially RCU bound in the high SNR region.
We also observe that the SGMC~$[128, 64]$ with the proposed NU profile outperforms that with the RM profile in the high SNR region, approaching the conventional RCU bound.

\end{example}

\subsection{Comparison with Finite-Length Bounds}

  \begin{figure}[tp]
    \centering
    \includegraphics[width=\figwidth]{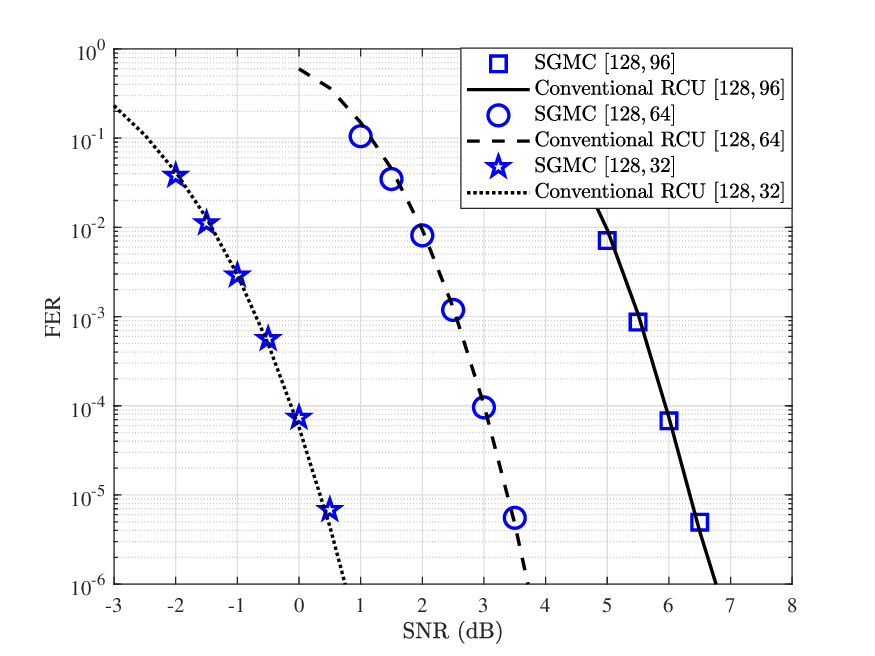}
    \caption{Performance of the proposed SGMCs with the NU profiles, where the LC-ROSD with $\delta = 12$ and $\ell_{\max} = 10^6$ is employed for decoding. The code length is $n =128$  and $w_0 = 50, 28,$ and $16$, respectively, for $k = 32,64$ and $96$.}
    \label{fig:ImprovedNU-FER-LB}
\end{figure}

\begin{example}
  Consider the SGMCs~(randomly generated but fixed) with the NU profiles, where the code length $n = 128$ and the dimension $k\in \{32,64,128\}$.
  The codewords are transmitted with BPSK modulation over the AWGN channels and the LC-ROSD is employed for decoding.
  The numerical results are shown in Fig.~\ref{fig:ImprovedNU-FER-LB}, from which we observe that the SGMCs with LC-ROSD can approach the corresponding conventional RCU bounds for different code rates.

\end{example}

\subsection{Comparison with 5G Codes}
\begin{figure}[tp]
    \centering
    \includegraphics[width=\figwidth]{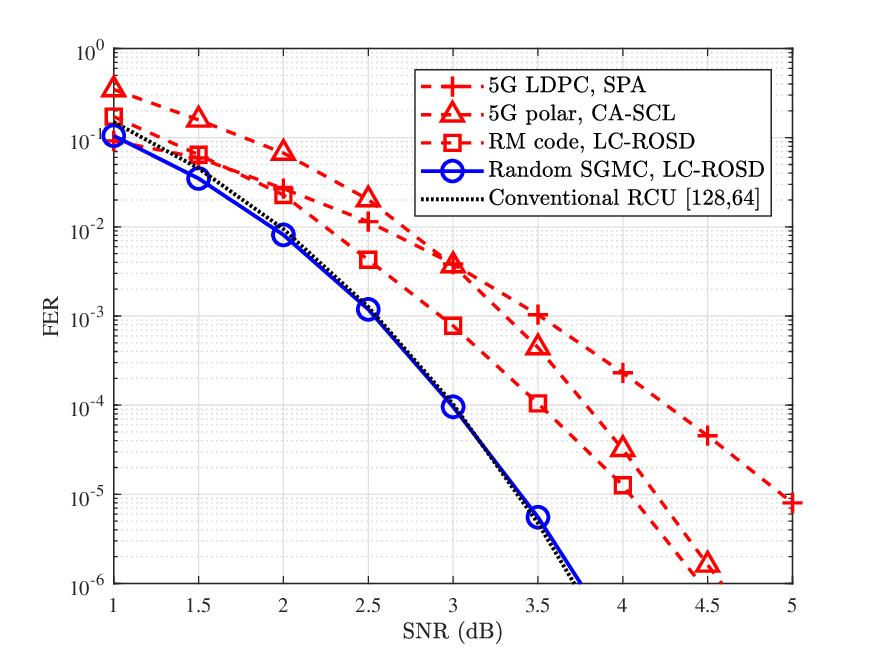}
    \caption{Performance of the proposed SGMC under the LC-ROSD~($\delta = 12$ and $\ell_{\max} = 10^6$), the RM code under the LC-ROSD~($\delta = 12$ and $\ell_{\max} = 10^6$), the 5G LDPC code under the SPA~(maximum iteration number of $50$) and the 5G polar code under the CA-SCL decoding~(list size of $16$). The code length is $n = 128$ and the dimension is $k = 64$.}
    \label{fig:Compare-5G-codes}
\end{figure}
\begin{example}
 Consider an SGMC~$\mathscr{C}[128,64]$~(randomly generated but fixed) with the NU profile, where the codewords are transmitted with BPSK modulation over the AWGN channels. For comparison, we also simulate the RM code under the LC-ROSD, the 5G LDPC code~\cite{5GeMBB} under the sum-product algorithm~(SPA)~(with a maximum iteration number of $50$) and the 5G polar code with 11-bit CRC~\cite{5GeMBB} under the CRC-aided successive cancellation list~(CA-SCL) decoding~(with a list size of $16$).
 The simulation results are shown in Fig.~\ref{fig:Compare-5G-codes}, from which we observe that the proposed SGMC can outperform these three codes, resulting in~(at ${\rm FER} = 10^{-5}$) a coding gain of up to $1.5$ dB over the 5G LDPC code, a coding gain of up to $0.7$ dB over the 5G polar code, and a coding gain of up to $0.6$ dB over to the RM code.
\end{example}

\subsection{Decoding Time Comparison with Other Decoding Algorithm}
\begin{figure}[tp]
    \centering
    \includegraphics[width=\figwidth]{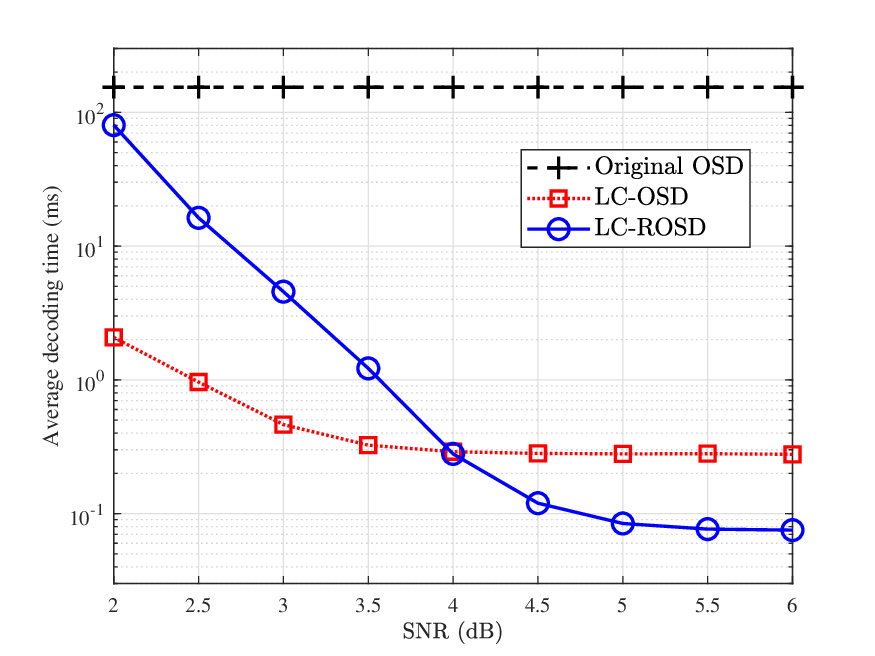}
    \caption{Average decoding time comparison for the proposed SGMC~$[128,64]$ with the NU profile, where we employ three decoding algorithms: the original OSD algorithm~(order $t=4$), the LC-OSD algorithm~($\delta = 12$ and $\ell_{\max} = 10^6$) and the LC-ROSD algorithm~($\delta = 12$ and $\ell_{\max} = 10^6$ ).}
    \label{fig:Time-Compare-other-OSD}
\end{figure}
\begin{example}\label{Example:decoding-time}
Consider an SGMC~$\mathscr{C}[128,64]$~(randomly generated but fixed) with the NU profile, where the codewords are transmitted with BPSK modulation over the AWGN channels. We have simulated three decoding algorithms: the original OSD~\cite{fossorier1995soft}, the LC-OSD~\cite{Liang2023LCOSD} and the LC-ROSD.
The average decoding time comparison is shown in Fig.~\ref{fig:Time-Compare-other-OSD}, from which we observe that, compared with the original OSD, which exhibits a constant decoding time, both the LC-ROSD and the LC-OSD exhibit a shorter decoding time~(on average).
In the low SNR region, where the re-encoding dominates the complexity, the average decoding time of the LC-ROSD is longer than that of the LC-OSD.
In contrast, in the high SNR region, the average decoding time of the LC-ROSD is shorter than that of the LC-OSD.
This speedup over the LC-OSD can be expected because the serial GE process becomes the main cause of the decoding latency due to the small number of TEPs in the high SNR region.

\end{example}

\section{Conclusions} \label{sec5}
This paper has proposed the random SGMCs, which have staircase-like generator matrices and enable parallel GE, resulting in a low-complexity LC-ROSD algorithm.
The random SGMCs have been demonstrated to be capacity-achieving over BIOS channels in the infinite-length region through the theoretical proof and have been demonstrated to be capacity-approaching in the finite-length region through the proposed tight bounds and the decoding performance.
The numerical results show that the proposed SGMCs with LC-ROSD match well with the proposed bounds.
The numerical results also show that the proposed random SGMCs outperform the 5G codes and RM codes.

\section*{Acknowledgment}
The first author would like to thank Dr. Suihua Cai and Xiangping Zheng for their helpful discussions.

\bibliographystyle{IEEEtran}
\bibliography{bibliofile}

\end{document}